\documentclass[prb,superscriptaddress,twocolumn,showpacs,amsmath,aps]{revtex4}
\usepackage{graphicx,color}
\usepackage{bm}
\usepackage[hypertex]{hyperref}

\def\tsigma{\tilde\sigma}

\begin{document}

\title{Does the measurement take place when nobody observes it?}

\author{Shmuel Gurvitz}
\email{shmuel.gurvitz@weizmann.ac.il}
\affiliation{Department of Particle Physics and Astrophysics\\
Weizmann Institute of
Science, Rehovot 76100, Israel}
\affiliation{ Beijing Computational Science Research Center,
Beijing 100084, China}
\date{\today}


\begin{abstract}
We consider {\em non-selective} continuous measurements of a particle tunneling to a reservoir of finite band-width ($\Lambda$). The particle is continuously monitored by frequent projective measurements (``quantum trajectory''), separated by a time-interval $\tau$. A simple analytical expression for the decay rate has been obtained. For Markovian reservoirs ($\Lambda\to\infty$), no effect of the measurements is found. Otherwise for a finite $\Lambda$, the decay rate always depends on the measurement time $\tau$. This result is compared with alternative calculations, with no intermediate measurements, but when the measurement device is included in the Schr\"odinger evolution. We found that the detector affects the system by the decoherence rate ($\Gamma_d$), related to the detector's signal. Although both treatments are different, the final results become very close for $\tau=2/\Gamma_d$. This $\tau$ corresponds to the minimal time for which the detector's signal can be distinguished by an ``observer''. This indicates a fundamental role of information in quantum motion and can be used for the extension of the quantum trajectory method for non-Markovian environments.
\end{abstract}

\maketitle
\section{Introduction}

\begin{figure}[h]
\includegraphics[width=8.5cm]{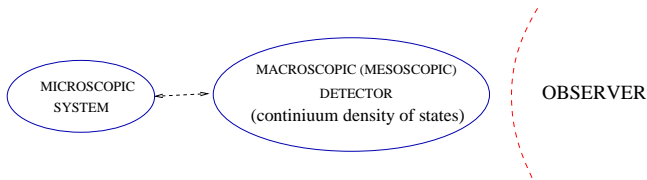}
\caption{(Color on line) Quantum Mechanical measurement. Microscopic system is coupled to another system (detector), perceive by an observer}
\label{fig11}
\end{figure}

A single quantum system is not observed directly, but through interaction with a macroscopic (mesoscopic) measurement device (detector) with continuous spectrum. The detector provides the measurement, which in quantum mechanics corresponds to projection of the {\em total} wave function including detector on eigenstates of the detector's variables, perceived by an external observer, Fig.~\ref{fig11}.
This is the so-called projection postulate\cite{neu}, which is analogues to the Bayes principle, inherent in any probabilistic description. Since the detector and the measured system are interacting, their variables are entangled. As a result, the projected total wave function can appear in an eigenstate (or in a superposition of eigenstates) of a variable belonging to the quantum system. The corresponding eigenvalue would represent the measurement's outcome.

This procedure implies a borderline between the ``detector'' and the ``observer'' (Fig.~\ref{fig11}), although the both are parts on the entire environment. This borderline  can be shifted towards the observer, by considering a chain of measurement devices representing the von Neumann hierarchy [1] (a system "measured" by another system etc). In order to avoid such uncertainty, one needs to assume that the shift of the borderline would not influence the  measurement outcome.

However a serious problem in this scheme arises in continuous measurements. Indeed, the projection postulate is usually applied at initial and final states  (preparation and measurement), whereas between the system undertakes the Schr\"odinger evolution. However, in the case of continuous measurement the detector remains switched-on all the time. Then there is no reason to apply the projection postulate only twice. One can imagine that detector is continuously monitored by an external observer, where all outcomes of such intermediate measurements are discarded (non-selective measurements). The question is whether such non-selective intermediate measurements can affect results of the final measurement?

At the first sight the {\em non-selective} intermediate measurements cannot influence the continuous evolution. Indeed each of such measurements (quantum jump) represents a sudden change in the observer's knowledge, not an objective physical event. This idea is implicit in the ``Quantum trajectory approach'', which considers quantum evolution of open systems as taking place under continuous observation \cite{WM10}. Nevertheless, repeated application of the projection postulate could affect quantum evolution of the total wave function, even for the non-selective measurements \cite{zeno}. It would be very desirable to investigate this problem on solvable realistic models for quantum system and detector. Then  we can explicitly compare the evolution of quantum system under frequent non-selective measurement with that given by the continuous Schr\"odinger evolution.

The plan of this paper is following. In Sec.~\ref{lab2} we consider a particle tunneling from a quantum well to a reservoir of a band-width $\Lambda$, monitored by an external observer with time intervals $\tau$. In order to determine how such repeated non-selective measurements influence the particle's  tunneling rate, we average over all possible quantum trajectories. In the case of finite $\Lambda$ we take into account the reversible dynamics, characterized tunneling to the non-Markovian reservoirs. Finally we arrived to a simple analytical expression for the tunneling rate, influence by continuous  measurements.

In Sec.~\ref{lab3} we consider continuous Schr\"odinger evolution of the system, when the observer is replaced by a point-contact (PC), monitoring the particle's  tunneling to the reservoir. The interaction with the PC affects the particle's tunneling rate. As in the previous case of continuous measurement, we obtain a simple analytical expression for the particle's tunneling rate. Comparing the both expressions, we find condition for the measurement time $\tau$, when the effect of the continuous measurement and the continuous Schr\"odinger evolution on the tunneling rate is the same. In Sec.~\ref{lab4} we discuss a possible meaning of this result and future investigations.

\section{Continuous non-selective measurements}\label{lab2}
\subsection{Markovian reservoir.}

\begin{figure}[h]
\includegraphics[width=5cm]{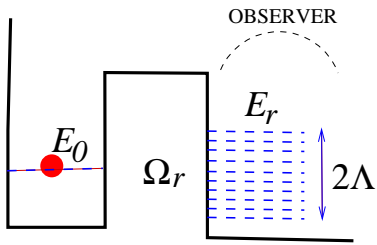}
\caption{(Color on line) Tunneling of a particle to continuum from a localized state inside the well to a reservoir of finite bandwidth $\Lambda$ The reservoir is monitored by an external observer.}
\label{fig1}
\end{figure}

Consider tunneling of a particle (electron) from a potential well (quantum dot) to a reservoir, Fig.~\ref{fig1}. The system is described by the following Hamiltonian
\begin{align}
H=E_0|0\rangle\langle 0|+\sum_rE_r|r\rangle\langle r|+\sum_r\Omega_r(|r\rangle\langle 0|+|0\rangle\langle r|)
\label{h1}
\end{align}
Here $|0\rangle$ is a localized state in the well and $|r\rangle$ denotes  extended states of the reservoir. The reservoir is monitored by an external observer, Fig.~\ref{fig1}.

The tunneling of a particle to the reservoir is described by the Schr\"odiger equation,
\begin{align}
i\partial_t |\Psi (t)\rangle =H|\Psi (t)\rangle
\label{sch}
\end{align}
where $|\Psi (t)\rangle$ can be written as
\begin{align}
|\Psi (t)\rangle=b_0(t)|0\rangle +\sum_rb_r(t)|r\rangle
\label{a3}
\end{align}
Here $b_{0}(t)$ is probability amplitude for finding the particle at the state $|0\rangle$ inside the well and $b_{r}(t)$ is the same for the state $|r\rangle$ inside the reservoir. Substituting Eq.~(\ref{a3}) into Eq.~(\ref{sch}) and performing the Laplace transform,
\begin{align}
\tilde b(E)=\int_0^\infty b(t)e^{iEt}dt,
\label{laplace}
\end{align}
where $E\to E+i0$ (causality condition), we can rewrite Eq.~(\ref{sch}) as
\begin{subequations}
\label{a4}
\begin{align}
&(E-E_0)\tilde b_0(E)-\sum_r\Omega_r\tilde b_r(E)= i\, b_0(0)\label{a4a}\\
&(E-E_r)\tilde b_r(E)-\Omega_r\tilde b_0(E)=i\, b_r(0)
\label{a4b}
\end{align}
\end{subequations}
where the right-hand-side corresponds to the initial conditions.

Consider the particle initially localized in the quantum well,  $b_0(0)=1$ and $b_r(0)=0$. Solving Eqs.~(\ref{a4}) in the continuous limit,  $\sum_r\to\int\rho(E_r)dE_r$, where $\rho(E_r)$ is the density of state, we find
\begin{align}
\tilde b_0(E)={i\over E-E_0-\int\limits_{-\infty}^{\infty} {\Omega^2(E_r)\rho(E_r)\over E-E_r}dE_r}
\label{m7}
\end{align}
with $\Omega_r^{}\equiv \Omega (E_r^{})$. In the case of Markovian reservoir (wide-band limit), the density of states and the coupling $\Omega (E_r^{})$ are independent of $E_r^{}$. Then integration over $E_r$ in Eq.~(\ref{mt}) can be easily performed, thus obtaining
\begin{align}
\tilde b_0(E)={i\over E-E_0+i{\Gamma\over2}}
\label{m71}
\end{align}
where $\Gamma =2\pi\Omega^2\rho$ (positive sign of $\Gamma$ is a result of $E\to E+i0$ in the integral (\ref{m7})). Note that for infinite reservoir, the density of states $\rho\sim L\to\infty$, where $L$ is the reservoir's size, but $\Omega^2\sim 1/L\to 0$, so the product $\Omega^2\rho$ (spectral density function) remains finite.

The amplitude $b_{0}(t)$ is obtained from $\tilde b_{0}(E)$ via the inverse Laplace transform,
\begin{align}
b_{0}(t)=\int\limits_{-\infty}^\infty \tilde b_{0}(E)e^{-iEt}{dE\over 2\pi}=e^{-iE_0t-{\Gamma\over2}t}
\label{invlap}
\end{align}
Thus in the wide-band limit, the particle initially localized inside the quantum well, decays exponentially to the reservoir.

Consider now the particle, initially localized at the level $E_{\bar r}^{}$ inside the reservoir: $b_0^{}(0)=0$ and $b_r^{}(0)=\delta_{r\bar r}^{}$ in Eqs.~(\ref{a4}). Then solving these equations for $b_0^{}(E)$ and using (\ref{invlap}), we find \cite{g0}
\begin{align}
b_0(t)={\Omega\over E_{\bar r}-E_0^{}+i{\Gamma\over2}}\Big(e^{-iE_{\bar r}^{}t}-e^{-iE_{0}^{}t-{\Gamma\over2}t}\Big)\to 0\, ,
\end{align}
since $\Omega\to 0$. This implies that for Markovian reservoirs, the particle detected in the reservoir at an extended state ($|\bar r\rangle$), will never appear inside the quantum well. The same can be shown if the particle is detected at a spatially localized state inside the reservoir, corresponding to a linear superposition of states $|\bar r\rangle$.

This above property of Markovian reservoir allows us to evaluate in a simple way the effect of continuous non-selective monitoring of the particle's tunneling to the reservoir. Indeed, if the particle is {\em not} detected in the reservoir, its new evolution takes place from the state $|0\rangle$ of the well. However, if the particle is detected in the reservoir, it never reappears inside the well. Therefore, such an event would not affect the particle decay from the well. Thus the probability of finding the particle inside the well
after a non-selective measurement of the reservoir at some time $t_1^{}$ is $P_0(t_1)=e^{-\Gamma t_1}$, Eq.~(\ref{invlap}). Then the probability of finding it in the well repeatedly, at $t>t_1$ is $e^{-\Gamma t_1}e^{-\Gamma (t-t_1)}=e^{-\Gamma t}$. It implies that the non-selective  monitoring of the Markovian reservoir does not lead to any observable effect in the tunneling  rate.

\subsection{Non-Markovian reservoir.}

Consider the reservoir of a finite band-width $\Lambda$, Fig.~\ref{fig1},  corresponding to the conduction band of periodic chain of quantum wells, with coupling $\lambda$, Eq.~(\ref{tbham1}) of Appendix, where the resulting spectral function is given by Eq.(\ref{tbham5}). We approximate this spectral density function of such a reservoir by a Lorentzian
\begin{align}
\Omega^2(E_r)\rho (E_r)={\Gamma\over 2\pi}{\Lambda^2\over (E_r-E_R)^2
+\Lambda^2}\, ,
\label{lor}
\end{align}
where $E_R$ is the Lorentzian center, and $\Lambda=2\sqrt{2}\lambda$, providing  the same curvature at the band-center that of Eq.(\ref{tbham5}). Although $|E_r^{}|$ in Eq.~(\ref{lor}) may exceed the bandwidth $\Lambda$, we show in Appendix that the Lorentzian (\ref{lor}) represents very good description for  finite band-width reservoir, Fig.~\ref{fig5}.

Substituting Eq.~(\ref{lor}) in Eq.~(\ref{m7}), we can evaluate the integral by closing the integration contour into lower complex $E_r$-plane. As a result, Eq.~(\ref{m7}) becomes
\begin{align}
(E-E_0)\tilde b_0(E)-{\Gamma\Lambda\over 2(E-E_R+i\Lambda) }\tilde b_0(E)=i
\label{a6}
\end{align}
(In the following we choose $E_0=0$.)
Using the inverse Laplace transform, Eq.~(\ref{invlap}), we obtain
\begin{align}
b_0^{}(t)=\, e^{-{Q\over2}t}
\Big[\cosh\Big({St\over2}\Big)
+{Q\over S}\sinh\Big({St\over2}\Big)\Big]
\label{m9}
\end{align}
where $Q=\Lambda+iE_R$ and $S=\sqrt{Q^2-2\Lambda\Gamma}$. In the limit $\Lambda\to\infty$ we return to the Markovian case by reproducing the exponential decay, Eq.~(\ref{invlap}), whereas for finite $\Lambda$, Eq.~(\ref{m9}) reproduces two-exponential decay. The difference with Markovian case is mostly significant for small times $t$. Indeed, the probability of decay to the Markovian reservoir, Eq.~(\ref{invlap}), reveals the irreversible dynamics, $1-|b_0(t)|^2_{}=\Gamma t+{\cal O}[t^2]$, whereas for the non-Markovian case, Eq,~(\ref{m9}) the dynamics is reversible,
\begin{align}
1-P_0(t)={\Gamma\Lambda\over2}t^2+{\cal O}[t^3]
\end{align}
As a result, the particle detected in the reservoir can reappear in the quantum well. This makes the treatment of non-selective measurements more involved, in comparison with the Markovian reservoir. For this reason we introduce a new basis for states of the non-Markovian reservoir, which can greatly simplify the problem.

\subsection{New basis of the reservoir's states.}

Consider Eq.~(\ref{a6}) for the amplitude $\tilde b_0(t)$. Let us introduce the  auxiliary amplitude (c.f. with Ref.~[\onlinecite{eg2}])
\begin{align}
\tilde b_R(E)={\bar\Omega\over E-E_R+i\Lambda}\tilde b_0(E)
\label{a7}
\end{align}
where
\begin{align}
\bar\Omega=\sqrt{{\Gamma\Lambda\over2}}
\label{ombar}
\end{align}
Then Eqs.~(\ref{a6}), (\ref{a7}) can be rewritten as
\begin{subequations}
\label{a8}
\begin{align}
&(E-E_0)\tilde b_0(E)-\bar\Omega \tilde b_R(E)=i\label{a8a}\\
&(E-E_R+i\Lambda )\tilde b_R(E)-\bar\Omega \tilde b_0(E)=0
\label{a8b}
\end{align}
\end{subequations}

\begin{figure}[h]
\includegraphics[width=6cm]{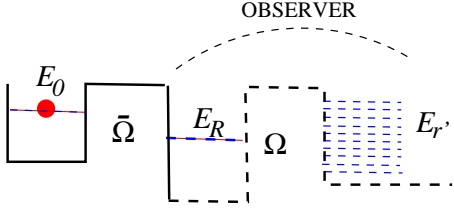}
\caption{(Color on line) Tunneling from the left well to continuum through the fictitious well. The level $E_R$ is at the Lorentzian center.}
\label{fig2}
\end{figure}

Let us demonstrate that Eqs.~(\ref{a8}) describe the particle in a  double-well, shown in Fig.~\ref{fig2}, where the second well is a fictitious one, coupled with a fictitious Markovian reservoir, with the coupling $\Omega$ and density of states $\rho$, such that  $\pi\Omega^2\rho=\Lambda$. This system is described by the Hamiltonian
\begin{align}
&H=E_0|0\rangle\langle 0|+E_R|R\rangle\langle R|
+\sum_{r'}E_{r'}|r'\rangle\langle r'|\nonumber\\
&+\bar\Omega(|R\rangle\langle 0|+|0\rangle\langle R|)
+\sum_{r'}\Omega(|r'\rangle\langle R|+|R\rangle\langle r'|)
\label{a1n}
\end{align}
Comparing (\ref{a1n}) with the original Hamiltonian, Eqs.~(\ref{h1}), we find that the reservoir's (extended) states $|r\rangle$ are split into the two components
\begin{align}
\sum_r|r\rangle\langle r|=|R\rangle\langle R|+\sum_{r'}|r'\rangle\langle r'|
\label{compn}
\end{align}
where $|r'\rangle$ represent the extended states of the fictitious {\em Markovian} reservoir. 

Now the particle wave function can be written in this new basis as
\begin{align}
|\Psi(t)\rangle=b_0(t)|0\rangle +b_R(t)|R\rangle+\sum_{r'}b_{r'}(t)|r'\rangle
\label{a3p}
\end{align}
Substituting it in the Schr\"odinger equation
$i\partial_t|\Psi(t)\rangle=H|\Psi(t)\rangle$ we find
\begin{subequations}
\label{ac4}
\begin{align}
&i\dot b_0(t)=E_0b_0(t)+\bar\Omega\, b_R(t)\label{ac4a}\\
&i\dot b_R(t)=E_Rb_R(t)+\bar\Omega b_0(t)+\sum_{r'}\Omega\, b_{r'}(t)
\label{ac4b}\\
&i\dot b_{r'}(t)=E_{r'}b_{r'}(t)+\Omega\, b_R(t)
\label{ac4c}
\end{align}
\end{subequations}
Resolving Eq.~(\ref{ac4c}) and substituting it into Eq.~(\ref{ac4b}), we obtain in the continuous limit, $\sum_{r'}\to \int\rho dE_{r'}$,
\begin{subequations}
\label{ab8}
\begin{align}
&i\dot b_0(t)=E_0 b_0(t)+\bar\Omega b_R(t)\label{ab8a}\\
&i\dot b_R(t)=(E_R-i\Lambda )b_R(t)+\bar\Omega\, b_0(t)
\label{ab8b}
\end{align}
\end{subequations}
After the Laplace transform, these equations coincide with Eqs.~(\ref{a8}).

\subsection{Continuous monitoring of the non-Markovian reservoir.}

Now we are going to evaluate the probability of tunneling to continuum under  the non-selective repeated  measurements by using a new basis (\ref{compn}) for the reservoir's states. Then if the particle  is detected at any of the states $|r'\rangle$, belonging to the fictitious Markovian reservoir, it cannot reappear inside the quantum or the fictitious wells. Therefore one needs to take into account only the states $|0\rangle$ and $|R\rangle$ in the corresponding quantum trajectories.

Note that the split of the reservoir basis into two components in no way implies a spacial separation of the non-Markovian reservoir into two parts. It remains the same reservoir of a finite band-width, like that discussed in Appendix. The composite of the fictitious middle well and Markovian reservoir in Fig.~\ref{fig2} still represents one system, which can be consider as a detector. The new basis is just a formal representation of the non-Markovian reservoir spectrum, which allows us to account the {\em reversible} dynamic in a most effective way.

Let us evaluate the amplitudes of finding the electron in the quantum or in the  fictitious well at time $\tau$, given by Eqs.~(\ref{ab8}). We denote these amplitudes as $b^{}_{0,R}(\tau )$ or $b'_{0,R}(\tau )$, for initial conditions corresponding to the occupied quantum or fictitious well:  $b_0^{}(0)=1$, $b_R^{}(0)=0$ or $b'_0(0)=0$, $b'_R(0)=1$. One finds (see Eq.~(\ref{m9}))
\begin{subequations}
\label{meas}
\begin{align}
&b_0^{}(\tau)=e^{-{Q\tau\over2}}\Big[\cosh\Big({S\tau\over2}\Big)
+{Q
\over S}\sinh\Big({S\tau\over2}\Big)\Big]\label{measa}\\[5pt]
&b_R^{}(\tau )=b'_0(\tau)=-i{\sqrt{2\Gamma\Lambda}\over  S}\,e^{-{Q\tau\over2}}\sinh\Big({S\tau\over2}\Big)\label{measb}
\\[5pt]
&b'_R(\tau)=e^{-{Q\tau\over2}}\Big[\cosh\Big({S\tau\over2}\Big)
-{Q\over S}\sinh\Big({S\tau\over2}\Big)\Big]
\label{measc}
\end{align}
\end{subequations}
The corresponding probabilities are denoted as $p_0=|b_0^{}(\tau)|^2$,  $p_1=|b_R^{}(\tau)|^2=|b'_0(\tau)|^2$ and $p_2=|b'_R(\tau)|^2$.

Now we can write the following recurrence relations for probabilities $P_{0,R}(m)$ to find the electron inside the quantum well or in the fictitious well after $m$ subsequent measurements, separated by the interval $\tau$, and starting from the initial condition $P_0^{}(0)=1$ and $P_R^{}(0)=0$,
\begin{align}
\left(\begin{array}{c}
P_0(m)\\P_R(m)\\
\end{array}
\right)=\left(
\begin{array}{cc}
p_0& p_1 \\
p_1&p_2\\
\end{array}
\right)
\left(\begin{array}{c}
P_0(m-1)\\P_R(m-1)\\
\end{array}
\right)
\label{pn}
\end{align}
Solving this equation we obtain
\begin{align}
\left(\begin{array}{c}
P_0(m)\\P_R(m)\\
\end{array}
\right)=\left(
\begin{array}{cc}
p_0& p_1 \\
p_1&p_2\\
\end{array}
\right)^m
\left(\begin{array}{c}
1\\0\\
\end{array}
\right)
\label{pn1}
\end{align}
As a result
\begin{align}
P_0(m)&={1\over2}\Big(1+{1\over\kappa}\Big)
\Big[{p_0\over2}(1+\kappa)+{p_2\over2}(1-\kappa)\Big]^m
\nonumber\\[5pt]
&+{1\over2}\Big(1-{1\over\kappa}\Big)
\Big[{p_0\over2}(1-\kappa)+{p_2\over2}(1+\kappa)\Big]^m
\label{cc}
\end{align}
where $\kappa=\sqrt{1+4p_1^2/(p_0-p_2)^2}$.

Finally the probability of finding the particle in its initial state at time $t$ after $m$ subsequent measurements is given by
\begin{align}
\sigma_{00}^{}(t)=P_0\left({t\over\tau}\right)
\label{cc1}
\end{align}
where $\tau=t/m$.

Consider the limit of continuous monitoring, $\tau\to 0$ ($m\to\infty$), by  assuming that $x=\Lambda \tau$ remains finite. By expanding $p_{0,1,2}^{}$ and $\kappa$ in powers of $1/m$, we find
\begin{align}
p_0&=1-{x+e^{-x}-1\over x\,m}\Gamma t+{\cal O}\left({1\over m^2}\right)\nonumber\\
p_1&={(1-e^{-x})^2\over 2mx}\Gamma t+{\cal O}\left({1\over m^2}\right)\nonumber\\
p_2&=e^{-2x}\Big(1+{1+x-e^x\over mx}\Gamma t\Big)+{\cal O}\left({1\over m^2}\right)\nonumber\\
\kappa&=1+{\cal O}\left({1\over m^2}\right)
\label{bet}
\end{align}

Substituting this result into Eq.~(\ref{cc}), we obtain
\begin{align}
P_0(m)=\left[1-\Big(1-{e^{-x}-1\over x}
\Big){\Gamma t\over m}+{\cal O}\left({1\over m^2}\right)\right]^m
\label{finprob3}
\end{align}
Then in the limit of continuous measurement, $m\to\infty$, we arrive to
\begin{align}
\sigma_{00}^{}(t)=e^{-\alpha \Gamma t}
\label{finprob1}
\end{align}
where
\begin{align}
\alpha\equiv\alpha(x)=1-{1-e^{-x}\over x}
\label{finprob2}
\end{align}
The same expression for the survival probability in limit of continuous measurement, has been obtained earlier for a particle transfer between two quantum wells through a reservoir, under the condition that no-particle is detected in the reservoir \cite{xq1} and also for the spontaneous photon emission \cite{xq2}.

It follows from Eq.~(\ref{finprob2}) that $\alpha(x)\to 1$ in the limit $x\to\infty$, corresponding to the Markovian reservoir. This implies no influence of measurement on the decay rate to continuum. However, for any finite $x$, the non-selective continuous measurement slows down the decay rate. It prevents it completely, $\alpha(x)\to 0$, in the limit of $x\to 0$ (quantum Zeno effect \cite{zeno}).

It is quite remarkable that a simple analytical expression for $\sigma_{00}^{}(t)$, Eq.~(\ref{finprob1}) reproduces the both Markovian and Zeno-effect limits (for $x\to\infty$ and $x\to 0$, respectively). This reflects that the measurement time ($\tau$) is not a most appropriate variable, for description of the continuous measurement, but the product of $x=\tau\Lambda$. Indeed, the limit of $\tau\to 0$ does not ensure the Zeno effect, but only $x\to 0$ (it follows from a simple explanation of the Zeno effect in terms of the energy-time uncertainty relation \cite{xq1}).

\begin{figure}[h]
\includegraphics[width=8.5cm]{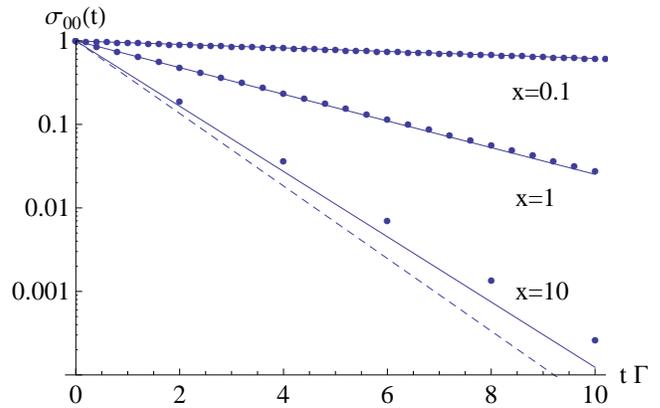}
\caption{(Color on line) Probability of finding the tunneling particle in its initial state as a function of time, under the non-selective measurements. The solid lines correspond to continuous measurements, whereas the dots show the results of discrete measurements for $\tau =x/\Lambda$ and $\Lambda=5\Gamma$. The dashed line shows the exponential decay, $\exp (-\Gamma t)$.}
\label{fig31}
\end{figure}

Although Eq.~(\ref{finprob1}) corresponds to the limit of continuous measurement, $\tau\to 0$ ($\Lambda =x/\tau\to\infty$), it reproduces very well the effect of measurements on $\sigma_{00}^{}(t)$ for finite $\tau$ and $\tau <t$, providing $\Lambda > \{\Gamma, E_R\}$. This is illustrated on Fig.~\ref{fig31} that displays $\sigma_{00}^{}(t)$ (in logarithmic scale), given by Eqs.~(\ref{finprob1}), (\ref{finprob2}) (solid lines) for $x=0.1$, $1$ and $10$, together with the result of Eqs.~(\ref{cc}), (\ref{cc1}) (dots) for $\Lambda=5\Gamma$ and $E_R=2 \Gamma$  shown by dots. For comparison we display the exponential decay $\sigma_{00}^{}(t)=\exp (-\Gamma t)$, undistorted by  measurements (dashed line).

We thus found that effect of non-selective measurements on decay to continuum, can be effectively accounted for by a simple factor $\alpha(x)$ that modifies the decay rate $\Gamma$  Eq.~(\ref{finprob2}), (\ref{finprob1}). This factor is explicitly dependent of the measurement time $\tau$, which has no fundamental meaning, since it is related to observer and does not appear in the Schr\"odinger equation of motion. Moreover, the outcome in non-selective measurements is discarded by the observer. Therefore the latter should play no role in the process and so the measurement time.

For understanding this problem, we should go to next level of the von Neumann hierarchy, where an observer is replaced by a device (detector), coupled to the tunneling electron. It would influences the electron tunneling rate, but now via their mutual interaction.

\section{Point-contact detector instead of observer}\label{lab3}
\subsection{General description.}

\begin{figure}[h]
\includegraphics[width=6cm]{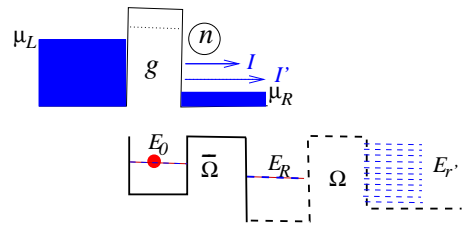}
\caption{(Color on line) Measurement of the quantum-well population with a point-contact detector. The current increases whenever the electron leaves the well. $n$ denotes the number of electrons arriving the right lead at time $t$.}
\label{fig3}
\end{figure}

Let us replace the observer in Figs.~\ref{fig1}, \ref{fig2} by a Point-Contact (PC) detector. The latter consists of two leads at different chemical potentials ($\mu_{L,R}^{}$) separated by a quantum PC and represented by a  potential barrier in Fig.~\ref{fig3}. If the PC is placed in close proximity to the quantum well, its opening decreases due repulsive electrostatic field of the electron, occupying the quantum dot. This results in increase of the barrier hight, and therefore in decrease on the electric current ($I$), flowing through the PC.  However, when the electron tunnels to the reservoir (to the fictitious well of Fig.~\ref{fig2}), its electric field near the PC decreases and the corresponding electric current increases, $I\to I'$ in Fig.~\ref{fig3}. Thus one can monitor the electron decay to continuum via the PC current.

In principle, we would have to put the point-contact detector near the reservoir, since it replaces the observer in Figs.~\ref{fig1}, \ref{fig2}. In practice, however, it is more difficult to realize, since the electron is delocalized in the reservoir, and therefore the detector needs to be much more sensitive. In addition, in the theoretical treatment the interaction of the PC with the reservoir would look more complicated. Therefore we put the PC near the dot. Then an increase of the PC current would imply detection of electron in the reservoir. In any case, the PC is not considered now as a measurement device, but as a part of the environment.

The entire system is described by the following Hamiltonian $H=H_0+H_{pc}+H_{int}$, where $H_0$ is given Eq.~(\ref{a1n}), $H_{pc}$ describes the PC detector, and $H_{int}$ is the interaction term. We use
\begin{eqnarray}
H_{PC}&=&\sum_l \bar E_l c_l^\dagger c_l+\sum_r \bar E_r c_r^\dagger c_r
+\sum_{l,r} [g\, c_l^\dagger c_r+H.c.]
\nonumber\\
H_{int}&=&\sum_{l,r} [\delta g\, c_l^\dagger c_r+H.c.]
|0\rangle\langle 0|\ ,
\label{bn1}
\end{eqnarray}
where the operators $c_{l(r)}^\dagger (c_{l(r)})$ corresponds to the creation
(annihilation) of electron in the state $\bar E_l(\bar E_r)$, belonging to the left (right) lead and $g$ is tunneling coupling
between these states. The quantity $\delta g=g'-g$
represents variation of the point contact hopping amplitude,
when the dot is occupied by the electron.

In contrast with the previous case of frequent non-selective measurements, the system undergoes continues Schr\"odinger evolution, $i\partial_t|\Psi (t)\rangle =H|\Psi (t)\rangle$, where $|\Psi (t)\rangle$ is the total wave many-particle function including the PC detector. The initial condition, $|\Psi (0)\rangle$, corresponds to the occupied quantum dot when the leads are filled up to the Fermi levels $\mu_L$ and $\mu_R$, Fig.~\ref{fig3}. The probability of finding the dot occupied at time time $t$ is $\sigma_{00}^{}(t)={\rm Tr}|\langle\Psi (t)|0\rangle|^2$, where the tracing takes place over all variables of the system. Solving the time-dependent Schr\"odinger equation one can evaluate $\sigma_{00}^{}(t)$, which is compared with that, Eq.~(\ref{cc1}), describing frequent non-selective measurements.

The problem can be solved analytically in the large bias limit, $V=\mu_L-\mu_R$. It was shown in Refs.~[\onlinecite{eg2,g1}] that in that limit the many-body Schr\"odinger  equation for $|\Psi (t)\rangle$ can by transformed  to master equations for the reduced density matrix
$\sigma_{jj'}^{(n)}(t)={\rm Tr}\langle j,n|\Psi (t)\rangle \langle\Psi (t)|j',n\rangle$, where $j(j')=\{ 0,R\}$, and $n$ is the number of electrons, arriving the right lead at time $t$, with $e n/t$ is the PC current. One finds
(see Eqs.(44a)-(44c) of Ref.~[\onlinecite{eg2}])
\begin{subequations}
\label{cn}
\begin{align}
&\dot{\sigma}_{00}^{(n)} = -I\sigma_{00}^{(n)}+I\sigma_{00}^{(n-1)}
+i\bar\Omega (\sigma_{0R}^{(n)}-\sigma_{R0}^{(n)})
\label{cn1}\\
&\dot{\sigma}_{RR}^{(n)} =  -(I'+2\Lambda )\sigma_{RR}^{(n)}
+I'\sigma_{RR}^{(n-1)}+i\bar\Omega (\sigma_{R0}^{(n)}-\sigma_{0RR}^{(n)})
\label{cn2}\\
&\dot{\sigma}_{0R}^{(n)} = i\epsilon\sigma_{0R}
+i\bar\Omega (\sigma_{00}^{(n)}-\sigma_{RR}^{(n)})
-\Big(\frac{I+I'}{2}+\Lambda\Big)\sigma_{0R}^{(n)}
\nonumber\\
&~~~~~~~~~~~~~~~~~~~~~~~~~~~~~~~~~~~~~~~~~
+\sqrt{I\, I'}\sigma_{0R}^{(n-1)}
\label{cn3}
\end{align}
\end{subequations}
where $\bar\Omega=\sqrt{\Gamma\Lambda /2}$, Eq.~(\ref{ombar}) and $\epsilon =E_0-E_R$. Here $I=2\pi g^2\rho_L\rho_RV$ is a current though the PC, when the quantum dot is occupied. Respectively, $I'=2\pi{g}^{\prime\, 2}\rho_L\rho_RV$ is is the same for the empty dot. (We use the units where the electron charge $e=1$).

The reduced density-matrix $\sigma_{jj'}^{(n)}(t)$ describes both the tunneling electron and the PC current. By tracing it over $n$, we find probability of the dot's occupation, $\sigma_{00}^{}(t)=\sum_n\sigma_{00}^{(n)}(t)$. Performing this procedure in Eqs.~(\ref{cn}) we obtain (c.f. with Eqs.(45a)-(45c) of Ref.~[\onlinecite{eg2}])
\begin{subequations}
\label{a11}
\begin{align}
&\dot\sigma_{00}=i\bar\Omega(\sigma_{0R}-\sigma_{R0})\label{a11a}\\
&\dot\sigma_{RR}=i\bar\Omega(\sigma_{R0}-\sigma_{0R})-2\Lambda
\sigma_{RR}\label{a11b}
\\
&\dot\sigma_{0R}=i\epsilon\sigma_{0R}+
i\bar\Omega(\sigma_{00}-\sigma_{RR})-\left({\Gamma_d\over2}+\Lambda\right)
\sigma_{0R}
\label{a11c}
\end{align}
\end{subequations}
where $\sigma_{jj'}^{}(t)=\sum_n\sigma_{jj'}^{(n)}(t)$ and
\begin{align}
\Gamma_d=(\sqrt{I}-\sqrt{I'})^2,
\label{decr}
\end{align}
These equations are of the Lindbladt (Bloch)-type Master equations and have a  clear physical meaning.  Indeed, in the case of no interaction with the PC detector ($I=I'$ and therefore $\Gamma_d=0$), one easily obtains Eqs.~(\ref{a11}) directly from Eqs.~(\ref{ab8}), taking into account that $\sigma_{jj'}^{}(t)=b_j^{}(t)b_{j'}^*(t)$. Hence, the interaction with the PC detector generates an additional damping rate ($\Gamma_d$) in Eq.~(\ref{a11c}) for the off-diagonal density-matrix element, $\sigma_{0R}(t)$. Since the latter is responsible for quantum-coherence effects in decay to continuum, we refer to $\Gamma_d$ as decoherence (dephasing) rate.

It order to solve Eqs.~(\ref{a11}) it is useful to apply Laplace transform, $\sigma (t)\to\tsigma(E)$, Eq.~(\ref{laplace}), thus obtaining
\begin{subequations}
\label{a13}
\begin{align}
&E\tsigma_{00}+\bar\Omega(\tsigma_{0R}-\tsigma_{R0})=i\label{a13a}\\
&(E+2i\Lambda )\tsigma_{RR}+\bar\Omega(\tsigma_{R0}-\tsigma_{0R})=0\label{a13b}
\\
&\left[E+\Delta+i\Lambda\left(1+{\Gamma_d\over 2\Lambda}\right)\right]\tsigma_{0R}+
\bar\Omega(\tsigma_{00}-\tsigma_{RR})=0
\label{a13c}
\end{align}
\end{subequations}
Solving Eqs.~(\ref{a13}) in the limit of $\Lambda\to\infty$, by keeping $\Gamma_d/\Lambda$ constant, we find
\begin{align}
\tsigma_{00} (E)=\frac{i \left(1+{2 \Lambda\over\Gamma_d}\right)}{E\left(1+{2\Lambda\over \Gamma_d}\right)+i\Gamma{2 \Lambda\over\Gamma_d}}
\label{a17}
\end{align}
Performing the inverse Laplace transform, Eq.~(\ref{invlap}),
by closing the contour of integration over the pole of $\tsigma_{00} (E)$, we finally obtain
\begin{align}
\sigma_{00}^{}(t)=e^{-\alpha'\Gamma t}
\label{scaling}
\end{align}
where
\begin{align}
\alpha'={2\Lambda \over\Gamma_d}\Big /\Big( 1+{2\Lambda\over\Gamma_d}\Big)
\label{scaling1}
\end{align}

\begin{figure}[h]
\includegraphics[width=6cm]{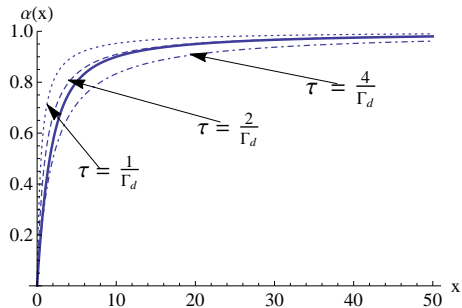}
\caption{(Color on line) Two exponential factors, $\alpha$ (solid line) and $\alpha'$ (dashed, dot-dashed and dotted lines) in Eqs.~(\ref{finprob1}), (\ref{scaling}) are displayed as a function of $x$.}
\label{mt}
\end{figure}

Let us compare Eqs.~(\ref{scaling}), (\ref{scaling1}) with Eqs.~(\ref{finprob1}), (\ref{finprob2}), obtained by repeated non-selective projective measurements. Although both expressions depend on different variables, they can be juxtaposed if the decoherence rate $\Gamma_d$ is inversely proportional to the measurement time (c.f. with  Refs.~[\onlinecite{shnir,kor}]), $\Gamma_d= c/\tau=c\Lambda/x$.
One finds that despite their different analytical forms, $\alpha(x)$ and $\alpha'(x)$   can be very close for certain values of $c$. For instance, for large $x$ it takes place for $c=2$, since $\alpha'(x)=\alpha (x)+{\cal O}(1/x^2)$ for $x\to\infty$. On the other hand, for small $x$ the best agreement between $\alpha$ and $\alpha'$ takes place for $c=4$, since then $\alpha'(x)=\alpha (x)+{\cal O}(x^2)$ for $x\to 0$. The result of such a comparison for different values of the coefficient $c$ is presented in Fig.~\ref{mt}, where $\alpha (x)$, Eq.~(\ref{finprob2}) is shown by solid line, and $\alpha' (x)$, Eq.~(\ref{scaling1}), is shown by dashed, dot-dashed and dotted lines, corresponding to $\tau =1/\Gamma_d,~2/\Gamma_d$ and $4/\Gamma_d$, respectively.

It is very remarkable that two different evaluations of $\sigma_{00}^{}(t)$, corresponding to repeated intermediate projections versus continuous evolution, can yield very close results for a  whole region of $x=\Lambda\tau$. Indeed, decoherence appears as a result of the continuous Schr\"odinger evolution, which excludes projective measurements at intermediate times. Hence, it would be natural to anticipate very different outcomes from the both treatments, as seemly displayed by Eqs.~(\ref{finprob2}), (\ref{scaling1}). It is therefore surprising that the results can be made very similar by assuming a simple relation between $\tau$ and decoherence rate, $\tau =c/\Gamma_d$. It is also remarkable that the best overall agreement is achieved for $c\simeq 2$. Indeed,  there are not fundamental restrictions for the coefficient $c$. For instance, nothing prevents to find it by orders of magnitude different from the above value. The question is whether this particular value of $c$ has a certain meaning. In order to understand this issues, we need to analyze behavior of the PC detector, as described by Eqs.~(\ref{cn}).

\subsection{PC current.}

Let us consider Eqs.~(\ref{cn}) for $\bar \Omega=0$, corresponding to a permanently occupied quantum dot. In this case the PC current is $I=n/t$, Fig.~\ref{fig3}, where its distribution of $n$ is given by Eq.~(\ref{cn1}). The latter now reads
\begin{align}
\dot P_n(t) = -I\big[P_n(t)-P_{n-1}(t)\big]\, ,
\label{pc1}
\end{align}
where $P_n(t)\equiv \sigma_{00}^{(n)}(t)$ is probability of finding $n$ electrons in the left lead. Note that the average number of electrons is $n(t)=\sum_nnP_n(t)$, so that  the electric current through the PC is $I(t)=\sum_nn\dot P_n(t)$. Using Eq.~(\ref{pc1}) one confirms that $I(t)=I$.

Solving Eq.~(\ref{pc1}) we find a Poisson distribution for $P_n(t)$ \cite{g2}
\begin{align}
P_n(t) = {(It)^n\over n!}e^{-It}\simeq {1\over\sqrt{2\pi It}}e^{-{(It-n)^2\over 2It}}
\label{pc2}
\end{align}
This distribution is centered around $n=It$ with a width $\sqrt{2It}$. Therefore the distribution's center displays the current flowing through the PC.

If the quantum dot is empty, the PC is described by the same Eq.~(\ref{pc1}), but  with $I\leftrightarrow I'$. In order to discern these two currents, the two distributions should be separated enough. One can define the minimal separation between two distributions, as such that sum of their widths equals to distance between their centers \cite{shnir}, $\sqrt{2It}+\sqrt{2I't}\le (I'-I)t$. This is equivalent to
\begin{align}
t\ge 2/(\sqrt{I'}-\sqrt{I})^2=2/\Gamma_d
\label{tm3}
\end{align}
where $\Gamma_d$ is the decoherence rate, Eq.~(\ref{decr}). The corresponding minimal time is therefore $t_{min}=2/\Gamma_d$.

\begin{figure}[h]
\includegraphics[width=0.22\textwidth,height=0.14\textheight]{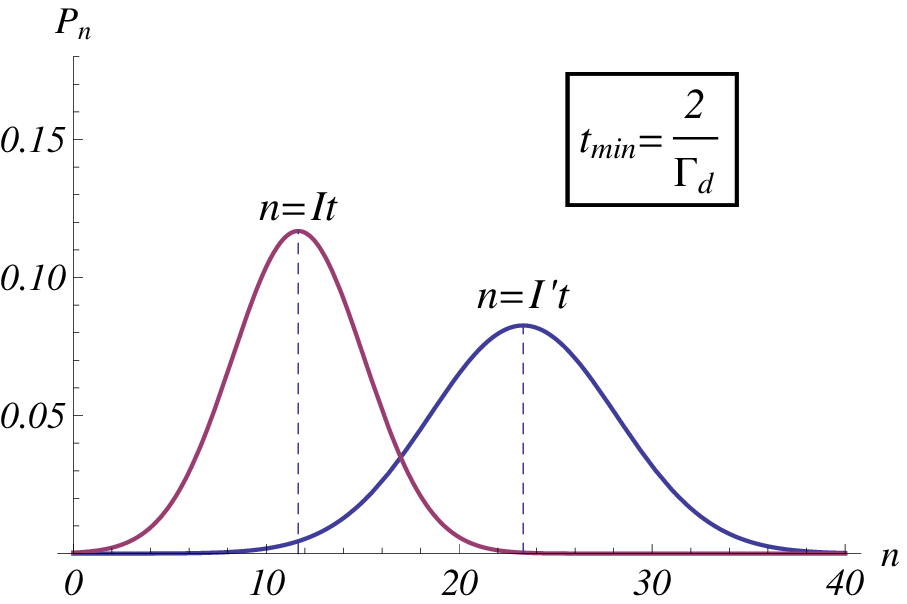}\ \ \
\includegraphics[width=0.22\textwidth,height=0.14\textheight]{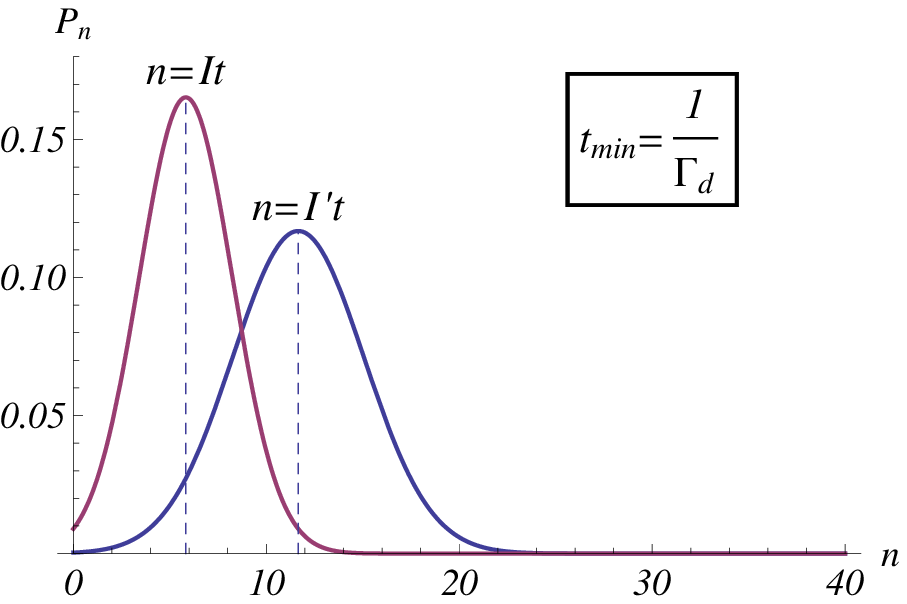}\ \
\caption{(Color on line) Distribution of charge $n$, arriving the right lead at time $t_{min}$ for two currents, $I=3$ and $I'=6$ (in arbitrary units). These currents  correspond to occupied and unoccupied quantum dot.}
\label{fig7}
\end{figure}

In fact, the minimal time, needed to discern these distributions is rather arbitrary quantity, since it is related to human perception. For instance, it can be defined as the minimal time when the signal-to-noise ratio is close to unity \cite{kor1}. This yields in $t_{min}= 1/\Gamma_d$. An example of two distributions, Eq.~(\ref{pc2}), for different $t_{min}$, is  shown in Fig.~\ref{fig7} for $I=3$ and $I'=6$ (in arbitrary units).

Thus the measurement time $\tau=2/\Gamma_d$, obtained from a comparison between continuous evolution versus continuous projections, corresponds to minimal time when the value of a ``signal'' equals to the width of its distribution. A possible interpretation of this result and its consequences are discussed in the next section.

\section{Discussion}\label{lab4}

In this work we study evolution of quantum system under repeated  non-selective projected measurements (quantum trajectories) versus continuous Schr\"odinger evolution. As an example we consider tunneling to non-Markovian reservoir, monitored by an external observer with time-interval $\tau$. We found that the repeated measurements affect the decay rate, even if the measurements outcome is not registered. At the next step, the observer is replaced by the PC detector that affects the decay rate, as well. However, in this case an entire system is described by continuous Schr\"odinger evolution with no intermediate projections. The question is whether the two different treatments would finally lead to the same result (modified decay rate)?

We demonstrate that this takes place if $\tau$ corresponds to minimal time when two detector's signals can be distinguished by an observer (the distance between two ``signal'' centers equals to sum of their width \cite{shnir}). Apparently, no such criterium of distinguishability enters the Schr\"odinger equation. Indeed, the quantum mechanics predicts only ensemble averaged quantities. Therefore if the ensemble is large enough, any signal can be distinguished no matter how weak it is. The above criterium is more related to human perception of noisy signals. Hence, it is very surprising that our analysis reveals $\tau$ in the "window" of human perception, Figs.~\ref{mt}, \ref{fig7}, which obviously has not been implicated in a comparison between between continuous measurement and Schr\"odinger evolution.

In fact, a similar comparison has been performed about 15 years ago in important work of Korotkov \cite{kor1}, which considered a qubit continuously monitored by a point-contact detector. Continuous measurement treated there with the Bayesian formalism \cite{kor}, has been confronted with  the ``Conventional approach'' (continuous Schr\"odinger evolution). The question asked there, ``Which equations are correct?'', received a final answer, namely ``all are correct depending on the problem considered.''

In the present paper we ask a similar question but in an opposite direction: at what condition the both approaches produce the same (or very close) output? This fix the measurement time $\tau$, which appears twice larger than than that used in the analysis of Ref.~[\onlinecite{kor1}]. The latter was taken as $1/\Gamma_d$ (in our notations), as the time for which the signal-to-noise ratio is close to unity. However, the simulations of Ref.~[\onlinecite{kor1}] have been less sensitive to the measurement time, in comparison with our analytical calculations of the continuous projections. We also concentrated on the non-Markovian dynamics, where the effect of the measurement time is more pronounced.

In fact, the detector is a part of the environment. Therefore in a more general sense, our result could imply that the environmental response, generated by the system, can be considered as a measurement, even in the absence of the observer. Such a measurement would always take place when the environmental signal can be distinguished by a ``potential observer'', using the above criterium of distinguishability. This point can also be considered as a definition of information through quantum dynamics.

Note that our result has been obtained in the framework of a particular setup. Therefore, it is necessary to investigate whether it is confirmed for different systems and detectors, like coupled quantum dots, or single-electron transistor instead of the PC. It is also very important to extend our analysis for the next levels of the von Neumann hierarchy. For instance, one can consider an observer, counting the number of electrons $n$ in the right lead, Fig.~\ref{fig3}, in comparison with the continuous evolution, when the observer is replaced by another device (``pointer''). Mostly important there would be the case when the lead is non-Markovian, in an analogy with Fig.~\ref{fig2}. Such a confirmation for universality of the measurement time can be important for understanding the quantum-classical transition. It also can be useful for extension of the quantum trajectory method on non-Markovian environment.

\begin{acknowledgements}
The author thanks Xin-Qi Li for useful discussions, and acknowledges the Beijing Normal University and the Beijing Computational Research Center for supporting his visit. This work was supported by the Israel Science Foundation
under grant No.\ 711091.
\end{acknowledgements}

\appendix
\section{Reservoir of finite band-width}\label{app1}

Reservoir of {\em finite} bandwidth $\Lambda$, Fig.~\ref{fig1}, corresponds to a periodic one-dimensional chain of $N$ quantum wells, with the nearest-neighbor coupling $\lambda$, describing  by the following Hamiltonian
\begin{align}
H_N=\sum_{n=1}^{N-1} \lambda(|n\rangle\langle n+1|+|n+1\rangle\langle n|)
\label{tbham1}
\end{align}
The state $|0\rangle$ of the quantum well is coupled with the first site of the chain by coupling $\tilde\Omega$, so the total Hamiltonian is $H=E_0|0\rangle\langle 0|+H_N+\tilde\Omega(|0\rangle\langle 1|+|1\rangle\langle 0|)$. By diagonalizing $H_N$ one arrives to Eq.~(\ref{h1}) with
\begin{subequations}
\label{tbham2}
\begin{align}
|r\rangle =&\sqrt{{2\over N+1}}\sum_{n=1}^N\sin \left({r\pi\over N+1}n\right)|n\rangle\label{tbham2a}\\
E_r^{}=&-2\lambda\cos \left({r\pi\over N+1}\right),~~{\rm for}~~ r=1,\ldots ,N\ ,
\label{tbham2b}
\end{align}
\end{subequations}
so that $-2\lambda<E_r<2\lambda$, and the corresponding spectral function is
\begin{align}
\Omega^2(E_r)\rho(E_r)={\Gamma\over2\pi}\sqrt{1-{E_r^2\over4\lambda^2}}
\label{tbham5}
\end{align}
where $\rho(E_r^{})=(dE_r/dr)^{-1}$ is the density of states and $\Gamma=\tilde\Omega^2/\lambda$. Here the band-center $E_R=0$.
The Markovian case (wide-band limit) corresponds to $\lambda\to\infty$. We assume that $\Gamma$ remains finite in this limit, which requires $\tilde\Omega\propto\sqrt{\lambda}\ll\lambda$.

\begin{figure}[h]
\includegraphics[width=7cm]{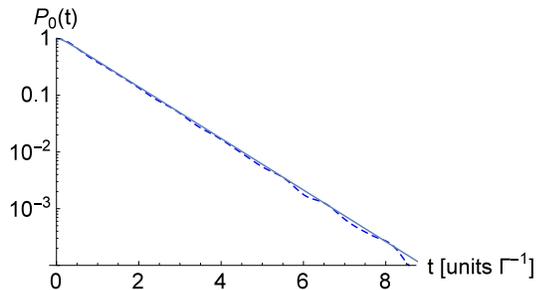}
\caption{(Color on line) Probability of finding the particle at its initial state at time $t$. Dashed line corresponds to periodic chain of $N=250$ coupled wells, and solid line shows continuous limit, $N\to\infty$, where the density of states is the Lorentzian, Eq.~(\ref{lor}).}
\label{fig5}
\end{figure}
In our calculations we approximate the spectral function (\ref{tbham5}) by the Lorentzian (\ref{lor}), since it allows us to treat the problem analytically without loosing its main physical features. For instance, Fig.~\ref{fig5} shows survival probability $P_0(t)=|b_0(t)|^2$, obtained from Eqs.~(\ref{a4}) and  (\ref{tbham2}) for $N=250$, $\lambda =3\Gamma$ and $E_0=\Gamma$ (dashed line) in comparison with the Lorentzian density of states, Eq.~(\ref{m9}), (solid line).  One finds that both curves almost coincide. This confirms that the Lorentzian (\ref{lor}) is a very good approximation for finite band-width reservoirs.

\end{document}